%


\magnification=\magstep0
\nopagenumbers
\parindent 25 pt
\parskip = 1 mm plus 0.3mm

\vsize = 23.80 cm
\hsize = 16.54 cm
\voffset = 0.0 mm
\hoffset = 0.0 mm
\tolerance=1200
\hyphenpenalty=1000
\newtoks\footline \footline={\hfil}
\vglue 0.05in

\def\sectionhead#1{\sectionskip\centerline{\twelvebf #1}\vskip6pt}
\def\sectionskip{\penalty-200\vskip24pt plus12pt minus6pt}

\def\eol     {\hfil\break}
\def\endpage {\vfill\supereject}


\medmuskip=4mu minus1mu
\thickmuskip=5mu minus1mu
\newcount\liczbamarg
\newcount\nrmarg
\newdimen\lmd
\newdimen\lml
\newdimen\ldd
\newdimen\ldl
\font\ogon=cmmi7
\def\ogonlow{\lower0.74ex}
\def\ogonleft{\kern-0.25em}
\lmd=0.05ex
\lml=0.31em
\ldd=0.12ex
\ldl=0.68em


\catcode`@\active
\def@{\leavevmode\futurelet\next\lettertest}
\def\lettertest
    {\ifx\next a\def\diacr{8}\else
     \ifx\next A\def\diacr{7}\else
     \ifx\next c\def\diacr{0}\else
     \ifx\next C\def\diacr{0}\else
     \ifx\next e\def\diacr{6}\else
     \ifx\next E\def\diacr{5}\else
     \ifx\next l\def\diacr{3}\else
     \ifx\next L\def\diacr{4}\else
     \ifx\next n\def\diacr{0}\else
     \ifx\next N\def\diacr{0}\else
     \ifx\next o\def\diacr{0}\else
     \ifx\next O\def\diacr{0}\else
     \ifx\next r\def\diacr{1}\else
     \ifx\next R\def\diacr{2}\else
     \ifx\next s\def\diacr{0}\else
     \ifx\next S\def\diacr{0}\else
     \ifx\next z\def\diacr{0}\else
     \ifx\next Z\def\diacr{0}\else
     \def\diacr{9}\char"40
     \fi\fi\fi\fi\fi
     \fi\fi\fi\fi\fi
     \fi\fi\fi\fi\fi
     \fi\fi\fi
     \ifcase\diacr
{\setbox0=\hbox{\next}\setbox1=\hbox{\accent"13\next}%
\ht1=\ht0\dp1=\dp0\box1}\or%
{\accent"5Fz}\or%
{\setbox0=\hbox{Z}\setbox1=\hbox{\accent"5FZ}\ht1=\ht0\dp1=\dp0\box1}\or%
{\char'40l}\or
{\setbox0=\hbox{L}\hbox to\wd0{\hss\char'40L}}\or
{\setbox0=\hbox{E}\setbox1=\hbox to \wd0{E\ogonleft\ogonlow%
  \hbox{\ogon\char"2C}\hss}\ht1=\ht0\dp1=\dp0\box1}\or%
{\setbox0=\hbox{e}\setbox1=\hbox to \wd0{e\ogonleft\ogonlow%
  \hbox{\ogon\char"2C}\hss}\ht1=\ht0\dp1=\dp0\box1}\or%
{\setbox0=\hbox{A}\setbox1=\hbox to \wd0{A\kern-0.24em\lower 0.72ex%
  \hbox{\ogon\char"2C}\hss}\ht1=\ht0\dp1=\dp0\box1}\or%
{\setbox0=\hbox{a}\setbox1=\hbox to \wd0{a\kern-0.21em\lower 0.75ex%
  \hbox{\ogon\char"2C}\hss}\ht1=\ht0\dp1=\dp0\box1}\or%
{\next}\fi%
\let\next}



\font\sixrm=cmr6
\font\sixi=cmmi6
\font\sixsy=cmsy6

\font\sevenrm=cmr7
\font\seveni=cmmi7
\font\sevensy=cmsy7

\font\tenrm=cmr10
\font\teni=cmmi10
\font\tensy=cmsy10
\font\tenit=cmti10
\font\tensl=cmsl10
\font\tenbf=cmbx10
\font\tentt=cmtt10

\font\twelverm=cmr10  scaled \magstep1
\font\twelvei=cmmi10  scaled \magstep1
\font\twelvesy=cmsy10 scaled \magstep1
\font\twelveit=cmti10 scaled \magstep1
\font\twelvesl=cmsl10 scaled \magstep1
\font\twelvebf=cmbx10 scaled \magstep1
\font\twelvett=cmtt10 scaled \magstep1

\font\fourteenbf=cmbx10 scaled \magstep2


\def\tenpoint{%
\def\rm{\fam0\tenrm}%
\def\it{\fam\itfam\tenit}%
\def\sl{\fam\slfam\tensl}%
\def\bf{\fam\bffam\tenbf}%
\def\tt{\fam\ttfam\tentt}%
\textfont0=\tenrm \scriptfont0=\sevenrm \scriptscriptfont0=\sixrm
\textfont1=\teni \scriptfont1=\seveni \scriptscriptfont1=\sixi
\textfont2=\tensy \scriptfont2=\sevensy \scriptscriptfont2=\sixsy
\textfont3=\tenex \scriptfont3=\tenex \scriptscriptfont3=\tenex
\textfont\itfam=\tenit \textfont\slfam=\tensl
\textfont\bffam=\tenbf \textfont\ttfam=\tentt
}

\def\twelvepoint{%
\def\rm{\fam0\twelverm}%
\def\it{\fam\itfam\twelveit}%
\def\sl{\fam\slfam\twelvesl}%
\def\bf{\fam\bffam\twelvebf}%
\def\tt{\fam\ttfam\twelvett}%
\def\cal{\twelvesy}%
 \textfont0=\twelverm \scriptfont0=\sevenrm \scriptscriptfont0=\sixrm
\textfont1=\twelvei \scriptfont1=\seveni \scriptscriptfont1=\sixi
\textfont2=\twelvesy \scriptfont2=\sevensy \scriptscriptfont2=\sixsy
\textfont3=\tenex \scriptfont3=\tenex \scriptscriptfont3=\tenex
\textfont\itfam=\twelveit \textfont\slfam=\twelvesl
\textfont\bffam=\twelvebf \textfont\ttfam=\twelvett
}


\raggedbottom		   
\def\deg{\ifmmode^\circ \else$^\circ$ \fi}	

\def\arcs{\ifmmode {'' }\else $'' $\fi}       
\def\arcm{\ifmmode {' }\else $' $\fi}         
\def\buildrel#1\over#2{\mathrel{\mathop{\null#2}\limits^{#1}}}
\def\mper{\ifmmode \buildrel m\over . \else $\buildrel m\over
      .$\fi}		       



\twelvepoint\rm \baselineskip 6.0 mm
\parindent 18 pt \parskip 0 mm

\input epsf.sty

\def\va{\vskip 5 mm}  \def\vc{\vskip 1.5 mm}
\def\ha{\hskip 5 mm}  \def\hc{\hskip 1.5 mm}

\centerline{\fourteenbf
Tables of the partition functions for iron, Fe I -- Fe X. }
\vskip 12 mm
\centerline{\bf J. Halenka$\,^1$, and J. Madej$^{\,2,\,3}$ }
\midinsert\obeylines \baselineskip 4.0 mm \leftskip 5 mm \vskip 12 mm
 $^1$Institute of Physics, University of Opole, Oleska 48, 45--052 Opole, Poland
\vc
 \ha\ha e-mail: halenka{@}uni.opole.pl

\vskip 5 mm
 $^2$Astronomical Observatory, University of Warsaw, Al. Ujazdowskie 4,
\vc
 \hc 00-478 Warszawa, Poland
\vc
 \ha\ha e-mail: jm{@}astrouw.edu.pl

\vskip 5 mm
 $^3$Copernicus Astronomical Center, Polish Academy of Sciences, Bartycka 18,
\vc
 \hc 00-716 Warszawa, Poland

\endinsert

\vskip 15 mm
\sectionhead{ABSTRACT}
\vskip 10 mm \baselineskip 6.0 mm
\newtoks\headline \headline={\hfil}

We present extensive tables of the Atomic Partition Function for
iron ions, Fe I -- Fe X, and discuss details of the computational
method. Partition functions are given in wide
range of temperatures, $10^3 \, {\rm K} < T < 10^6$ K, and lowering
of ionization energy ($0.001 \, {\rm eV}< LIE < 5.0 $ eV).
Our $APF$ take into account all energy
levels predicted by quantum mechanics, including autoionization levels.
The tables can be applied for the computations of model
stellar atmospheres and theoretical spectra over all existing
spectral and luminosity classes.

Our tables are available at
{\tt http://www.astrouw.edu.pl/$\sim$acta/acta.html}
(Acta Astronomica Archive),
and {\tt http://draco.uni.opole.pl/Halenka.html}.

\vskip 2 cm
\line{{\bf Key words:} Atomic data -- Plasmas -- Stars: atmospheres \hfil}

\endpage

\line{}
\sectionhead{1. Introduction}
\vskip 3 mm

Iron is the most abundant and important heavy element in
astrophysical plasmas. Due to its relatively high abundance in solar
type composition and low ionization energy, iron ions are important
contributors of free electrons in stellar atmospheres of population II
at low and moderate effective temperatures (spectral types F and lower).
In atmospheres of these types hydrogen and helium atoms remain almost
perfectly neutral, and therefore iron donors of free electrons can be
critically important for the populations and opacity of negative ions
H$^{-}$ and He$^{-}$, which in turn determine the whole atmospheric
structure and visual/infrared spectra of these most numerous classes
of stars.
\newtoks\headline \headline={\hss\tenrm\folio\hss}

Also on the opposite end of the Hertzsprung-Russell diagram, various ions
of iron contribute to the stratification of monochromatic bound-free and
line opacities in stellar atmospheres of the highest effective temperatures,
including both main sequence and white dwarf stars. Highly ionized iron ions
efficiently absorb extreme ultraviolet and X-ray photons, where the outgoing
flux of radiation is most intense in these objects. Therefore iron plays a
crucial role in determining the structure of atmospheres and appearance of
the radiation spectra of the hottest stars.

Both the above examples show the importance of construction of the
partition functions for various iron ions, which are internally
homogeneous and can determine exactly populations of iron ions over very
wide range of temperatures and electron concentrations. Our paper fulfils
such a demand.

Values of the partition function $U$ for a particular ion depend
on gas temperature $T$, and the local electron concentration $N_e$.
Temperature enters partition function by its definition (Griem 1964,
Drawin and Felenbok 1965, Traving et al. 1966)

$$ U (T, N_e) = \sum\limits^{i_{max}}_{i=1} g_i \exp (-E_i / kT) \, ,
   \eqno(1) $$

\noindent
where the sum is taken over discrete energy levels of statistical weights
$i$ and excitation energies $E_i$. In vacuum ($N_e = 0$) the number of
bound levels $i_{max}$ is infinite, and therefore the above series diverges.

In real plasma, however, interaction between the atom of interest and
surrounding free electrons and ions cause, that bound
levels of very high excitation energies move to continuum and no longer
contribute to the partition function. Therefore the series in Eq. (1)
reduces to finite number of terms, and value of $U$ is also finite
and is strongly dependent on the electron concentration $N_e$. In
general, the larger is $N_e$ the lower is both number of bound energy
levels $i_{max}$ in Eq. (1), and the value of $U$.

There exist a large number of papers, which present tables of rather
approximate (or even schematic) partition functions ($APF$) for elements,
including few iron ions (Drawin and Felenbok 1965; Traving et al. 1966;
Irwin 1981, for example). The latter paper presents fitting formulae for
$APF$ of Fe I -- Fe III, for temperatures $T \le 16000$ K. The widely
used computer code Tlusty 195 for computations of NLTE model stellar
atmospheres (Hubeny and Lanz 1992, 1995) contains FORTRAN subroutine
computing $APF$ of Fe IV -- Fe IX by direct summation over all
energy levels {\sl observed} in these ions in 1971. Therefore the set
of energy levels was not complete in TLUSTY195.

Quality of given tables of partition functions depends on ($i$)
accuracy and completeness of energy levels included in Eq. (1),
and ($ii$) realiability of the assumed theory of emitter-plasma
interactions. Emitter-plasma interactions cause that the series in
Eq. (1) is finite. Unfortunately, none of the currently existing theories
describe correctly effects of charged particles in plasma on the
atomic partition functions, cf. also Hummer and Mihalas (1988).

We have computed and present tables
of the partition functions taking into account all energy
levels predicted by quantum mechanics, including also levels lying
above the so called {\sl normal ionization energy} (autoionization levels).
Moreover, our computations are more physically correct, since values of
$APF$ depend on both temperature $T$ and electron concentration $N_e$.
The latter is related to the lowering of ionization energy ($LIE$).
The method of $APF$ calculations is described in the following
Section. This method was also presented in series
of papers by Halenka and Grabowski (1977, 1984, 1986), Halenka (1988, 1989),
and Madej et al. (1999). Paper by Halenka et al. (2001, hereafter Paper I)
presented already extensive tables of $APF$ for nickel, Ni I -- Ni X, which
were computed in the same way.

\sectionhead{2. Computations of the atomic partition functions}
\vskip 3 mm

We have appended contribution to the Atomic Partition Function given
by autoionizing levels. We replace the classical definition of $APF$,
Eq. (1), by more realistic expression introduced by Halenka and Grabowski
(1977) as follows

$$ U^ {(r)} (T, N_e) = \sum\limits_{p=1}^{p_{max}}
   \sum\limits_{i=1}^{i(p)_{max}} g^{(r)}_{pi} \exp(-E^{(r)}_{pi} /kT)
   = \sum\limits_{p=1}^{p_{max}} U^{(r)}_p (T, N_e) \, ,   \eqno(2) $$

\noindent
Here the set ($pi$) with the indices $p$ and $i$ (ordering of levels
from the ground level upwards in energy scale) describes an eigenstate
of the atom in the $r$-th ionization state. Index $i$ represents three
quantum numbers ($nlj$) of the optical electron, and $p$ represents the
quantum state of the atomic core. Index $i(p)_{max}$ is the number of
all bound energy levels, $g^{(r)}_{pi}$ and $E^{(r)}_{pi}$ denote
statistical weight and excitation energy of the $i$-th state, in the
sequence based on the $p$-th parent level. Numbers $i(p)_{max}$ result
from the inequality

$$ E^{(r)}_{pi} \le E^{(r)}_{p\infty} - \Delta E^{(r)} \, ,  \eqno(3) $$

\noindent
where $\Delta E^{(r)}$ denotes the lowering of the ionization energy $LIE$,
and $E^{(r)}_{p\infty}$ is the ionization energy in the $p$-th level
sequence. The latter quantity is equal to the sum

$$ E^{(r)}_{p\infty} = E^{(r)}_{1\infty} + E^{(r+1)}_p \, .  \eqno(4)$$

\noindent
The quantity $ E^{(r+1)}_p$ denotes the energy of the atomic core after
ionization, $r \rightarrow r+1$.
Index $p_{max}$ is the number of different parent levels which can
be realized in given physical conditions. The number $p_{max}$ results
from the inequality similar to Eq. (3), written for the $(r+1)$-th
ionization state. Since for a fixed value of $p$ the number $k$ is
assigned unambiguously, then the upper limit of $E^{(r)}_{p\infty}$ for
the $k$-fold excitation can be written as follows. Numbers $k=1,2, \ldots,
Z-r$, where $Z$ denotes the atomic number

$$ E^{(r)}_{p\infty} \le \sum\limits^k_{s=1} E^{(r+s-1)}_{1\infty} \, .
   \eqno(5) $$

The above definition of $APF$ requires, that for a fixed ionization
state $r$ the initial set of observed and calculated energy levels must
be split into level sequences according to their attachment to different
parent terms. Our initial set of levels was taken from catalogue of
energy levels of iron ions from the CD-ROM by Kurucz (1994), which contained
both the 'observed' and computed energy levels (of the order 1000 for each
ionization state $r$). We have also added to our $APF$'s contribution from
many energy levels predicted by quantum mechanics, which were  missed in
CD-ROM by Kurucz. Exact description of the method of adding of missing
levels was given by Halenka and Grabowski (1977, 1984).

Given term of a configuration with equivalent electrons usually belongs
simultaneously to few level sequences defined by different parent terms.
To calculate the contribution of a fixed term $i$
to the partial $APF$, $U^{(r)}_{p}$ in Eq.~(2),
one must multiply the statistical weights of the $i$-th term
by their coefficients of the fractional parentage $G_{qi}$, which were
normalized to $1$.
We have assigned coefficients of the fractional parentage according to
the well known procedure (cf. Slater 1960, vol. II; Sobelman 1979):

$$ G_{pi} \equiv G_{\alpha^{'}L^{'}S^{'}; {\alpha}LS } =
   (\alpha^{'}L^{'}S^{'}; {\alpha}LS)^2/
    \sum{}^{'} \alpha^{'}L^{'}S^{'}; {\alpha}LS)^2.
    \eqno(6)$$

For a fixed ionization degree $r$, all level sequences were divided
into three groups: $\alpha^{(r)}$, $\beta^{(r)}$, and $\gamma^{(r)}$.

Group $\alpha^{(r)}$ consists of level sequences numbered
by $ p=1,2,...p_1$, where all energy levels (energies and statistical
weights) are known. Usually the total number of such level sequences
does not exceed 20.
In most cases these level sequences belong to the ground parent
configuration. Their contribution to the $APF$ is given by
summation of terms in Eq.~(2).

Group $\beta^{(r)}$ consists of level sequences, in which only part of
energy levels are known, from $i=1$ up  $i_{1}(p)$.
Number of such level sequences, $p=p_1+1,\ldots, p_2$, never exceeds one
hundred.
Contributions of $\beta^{(r)}$ level sequences to the $APF$
is given by summation up to level $i_{1}(p)$. Remaining levels are
included approximately, assuming their similarity to some of level
sequence of the group $\alpha^{(r)}$, so called reference level sequence.
Thus, for $p_1<p\leq p_2$, values of $U^{(r)}_{p}$ can be calculated
according to the relation:

$$
\sum\limits^{}_{p \in term~s} U^{(r)}_{p} = \sum\limits^{}_{p \in term~s}
   \sum\limits^{i_{1}(p)}_{i=1} g^{(r)}_{pi}
   \exp(-E^{(r)}_{pi}/kT)+ \sum\limits^{}_{p \in term~s}\Delta
   U^{(r)}_{p} \, , \eqno(7)$$

\noindent where
$$
\sum\limits^{}_{p \in term~s} \Delta U^{(r)}_{p} \simeq
\sum\limits^{}_{p \in term~q}\Delta U^{(r)}_{p}
{{
\sum\limits^{}_{p \in term~s} g^{(r+1)}_{p} \exp(-E^{(r+1)}_{pi}/kT)}
\over {\sum\limits^{}_{p \in term~q} g^{(r+1)}_{p}
\exp(-E^{(r+1)}_{pi}/kT)} } \, ,
   \eqno(7')
$$
\noindent and
$$ g_{pi} = (2J_{pi} +1)G_{pi}\, . $$
Here $q$ ($q \le p_1$) represents a reference level sequence.

 Group $\gamma^{(r)}$ is the set of level sequences, in
which no levels are known. Their share in the $APF$ can be estimated
from the expression

$$ \sum\limits^{p_{max}}_{p_2+1} U^{(r)}_{p} =
   { \sum\limits^{}_{p \in term~q} U^{(r)}_{p} \over
   {\sum\limits^{}_{p \in term~q} g^{(r+1)}_{p} \exp(-E^{(r+1)}_{pi}/kT)}}
    \times \left[ U^{(r+1)} - \sum\limits^{p_2}_{p=1}
     g^{(r+1)}_{p} \exp(-E^{(r+1)}_{pi}/kT) \right] \, .
   \eqno(8)
$$

\noindent
The relation Eq.~(7') is strictly valid if, at a
fixed quantum number $J$ of the optical electron, change $s \rightarrow q$
of the state of the atomic core does not alter the electron-core interaction
energy. In such a case energy levels of the $s$ sequence are shifted as a whole
relatively to $q$ sequence by energy $|E^{(r+1)}_{s} - E^{(r+1)}_{q}|$,
and  the statistical weighs satisfy the relation
$g^{(r)}_{s}/g^{(r)}_{q} = g^{(r+1)}_{s}/g^{(r+1)}_{s}$.

The above condition is never strictly realized. However, the error introduced
into $U^{(r)}_{p}$ by the Eq.~(7') reaches minimum when the maximum bound
energies in both considered  level sequences ($s$ and $q$)
reach similar values.
It should be noted that values of $U^{(r)}$ are of physical interest
only in such range of temperatures $T$, when $E^{(r)}_{p\infty}/kT \gg 1$.
Since for $p>p_1$ we have
$E^{(r)}_{pi} \geq E^{(r)}_{1\infty}$ (except for few of the lowest
levels), then the error contributed by $\Delta U^{(r)}_{p}$ to $U^{(e)}$
in Eq.~(7') is always small.

\sectionhead{3. Results}
\vskip 3 mm

We have computed $APF$ for Fe~I - Fe~X ions according to above procedure.
Exemplary results shown in Fig. 1, which  displays run of our
$APF$ for Fe IV and temperatures corresponding to atmospheres of hot white
dwarf stars, for various values of $LIE$ taken as free parameter (solid
lines). Numerical values of log $APF$ for Fe IV are also listed in Table 1.
In Fig. 1 we have appended run of partition functions computed on the
basis of the {\sl observed} levels only (single dashed line). One can
easily note, that our recommended partition functions are significantly
larger than the latter functions, and apparently tend to diverge at any
fixed $T$ and $N_e \rightarrow 0$.

Partition functions for iron are arranged in 10 ASCII tables, where
each table corresponds to a single ion, which is identical with the
organisation of Ni I -- Ni X partition functions of Paper I.
Full set of these data is available from Acta Astronomica Archive, \eol
{\tt http://www.astrouw.edu.pl/$\sim$acta/acta.html}, \eol
or {\tt http://draco.uni.opole.pl/Halenka.html}.

Entries of tables are decimal
logarithms of the APF. They are tabulated at 62 discrete temperatures,
spaced at nonequidistant intervals, $10^3 \le T \le 10^6$ K, and at
9 arbitrarily assumed values of the lowering of ionization energy ($LIE$
= 0.001, 0.003, 0.010, 0.030, 0.100, 0.300, 1.000, 3.000, and 5.000 eV).
Both $T$ and $LIE$ points remain identical in all 10 tables of APF, to
ensure homogeneity of the data and compatibility with Paper I.

Very wide range of both parameters $T$ and $LIE$ assumed in our tables
ensures, that they cover practically all the conditions
expected in stellar atmospheres of any type, reaching
from the coldest dwarfs up to the hottest accreting white dwarf stars.

Tables discussed in this paper are based on the most complete set of
energy levels actually available for iron ions. Moreover, our partition
functions are sensitive to plasma interactions, since they strongly depend
on the lowering of ionization energy. Our $APF$
tend to diverge for the lowering of ionization energy approaching zero,
and are perfectly compatible with the $APF$ for nickel ions (cf. Paper I).

\epsfxsize=15.0cm \epsfbox[60 80 630 575]{xac3.fig}

\leftskip 19 mm \rightskip 31 mm
\tenpoint\rm \baselineskip 4.5 mm

\vskip -15 mm
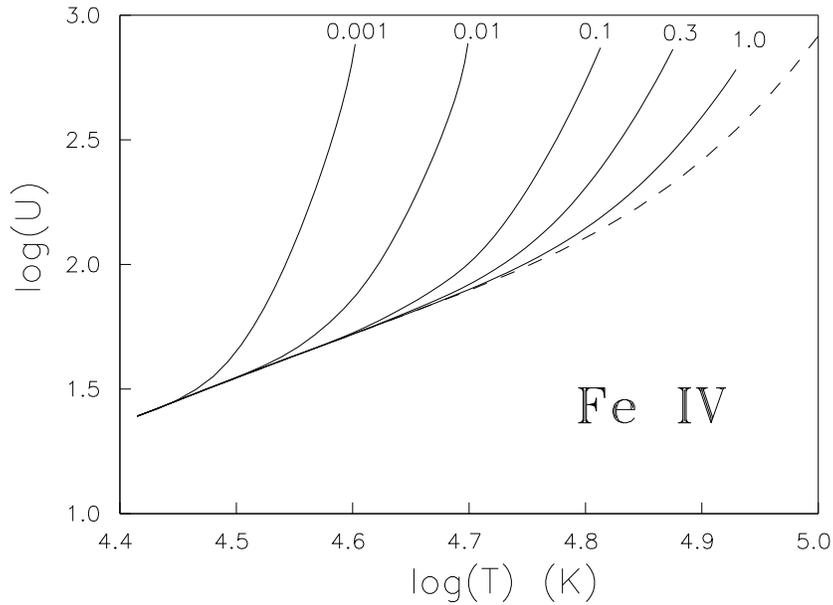
{\bf Fig. 1 :} Run of partition functions of Fe IV as function of gas
temperature $T$ and various parameters $LIE$. Solid lines represent our
recommended results, whereas dashed line represents partition function
computed from the observed levels only.

\twelvepoint\rm \baselineskip 6.0 mm
\parindent 21 pt \parskip 0 mm
\leftskip 0 mm \rightskip 0 mm

\vskip 20 mm
\line{\bf Acknowledgements \hfil} \va
We are grateful to R.L. Kurucz for making his CD-ROM No. 22 available
for us. JH and JM acknowledge support by grant No. 2 P03D 021 22 from the
Polish Committee for Scientific Research.

\endpage

\line{}
\sectionhead{REFERENCES}
\va
\dimen0=\hsize  \advance\dimen0 by -25 pt
\def\ref#1{\parshape=2  0.pt \hsize  25pt \dimen0 #1}

\parindent 0pt
\parskip 8pt

\ref Drawin, H.W., and Felenbok, P. 1965, Data for Plasma in Local
     Thermodynamic Equilibrium (Gauthier-Villars, Paris)

\ref Griem, H.R. 1964, Plasma Spectroscopy (McGraw-Hill Book Co.,
     New York)

\ref Halenka, J. 1988, Astron. Astrophys. Suppl., {\bf 75}, 47

\ref Halenka, J. 1989, Astron. Astrophys. Suppl., {\bf 81}, 303

\ref Halenka, J., and Grabowski, B. 1977, Astron. Astrophys., {\bf 54}, 757

\ref Halenka, J., and Grabowski, B. 1984, Astron. Astrophys. Suppl.,
     {\bf 57}, 43

\ref Halenka, J., and Grabowski, B. 1986, Astron. Astrophys. Suppl.,
     {\bf 64}, 495

\ref Halenka, J., Madej, J., Langer, K., and Mamok, A. 2001,
     Acta Astron., {\bf 51}, 347 (Paper I)

\ref Hubeny, I., and Lanz, T. 1992, Astron. Astrophys., {\bf 262}, 501

\ref Hubeny, I., and Lanz, T. 1995, Astrophys. J, {\bf 439}, 875

\ref Hummer, D.G., and Mihalas, D. 1988, Astrophys. J., {\bf 331}, 794

\ref Irwin, A.W. 1981, Astrophys. J. Suppl., {\bf 45}, 621

\ref Kurucz, R.L. 1994, CD-ROM No. 22

\ref Madej, J., Halenka, J., and Grabowski, B. 1999, Astron. Astrophys.,
     {\bf 343}, 531

\ref Mihalas, D. 1978, Stellar Atmospheres (Freeman, San Francisco)

\ref Slater, J.C. 1960, Quantum Theory of Atomic Structure, vol. II
     (McGraw - Hill Book Co., New York)

\ref Sobelman, I.I. 1979, Atomic Spectra and Radiative Transitions
     (Springer - Verlag, Berlin)

\ref Traving, G., Baschek, B., and Holweger, H. 1966, Abh. Hamburger
    Sternw., {\bf VIII}, No. 1

\parindent 21 pt
\endpage

\tenpoint\rm \baselineskip 3.5 mm

\newdimen\digitwidth
\setbox0=\hbox{\rm0}
\digitwidth=\wd0
\catcode`?=\active
\def?{\kern\digitwidth}
\def\-{$-$}

\newtoks\headline \headline={\hfil}

\halign {\hfil#\hfil& \hskip 3 mm \hfil#\quad\hfil& \hfil#\quad\hfil
    & \hfil#\quad\hfil & \hfil#\quad\hfil & \hfil#\quad\hfil &
\hfil#\quad\hfil
    & \hfil#\quad\hfil & \hfil#\quad\hfil & \hfil#\quad\hfil   \cr
\noalign{\centerline{Table 1} \smallskip}
\noalign{\centerline{Atomic partition functions for Fe IV \hc
   (decimal logarithms)} }
\noalign{\medskip \hrule \smallskip }
\noalign{\centerline{Lowering of ionization energy (eV)}  \smallskip }
  \ha T (K)& ??0.001& ??0.003& ??0.010& ??0.030& ??0.100& ??0.300&
             ??1.000& ??3.000& ??5.000  \cr
\noalign{\smallskip \hrule \smallskip}
???1000&??.7782&??.7782&??.7782&??.7782&??.7782&??.7782&??.7782&??.7782&??.7782\cr
???3000&??.7782&??.7782&??.7782&??.7782&??.7782&??.7782&??.7782&??.7782&??.7782\cr
???6000&??.7800&??.7800&??.7800&??.7800&??.7800&??.7800&??.7800&??.7800&??.7800\cr
???8000&??.7905&??.7905&??.7905&??.7905&??.7905&??.7905&??.7905&??.7905&??.7905\cr
??10000&??.8196&??.8196&??.8196&??.8196&??.8196&??.8196&??.8196&??.8196&??.8196\cr
??12000&??.8709&??.8709&??.8709&??.8709&??.8709&??.8709&??.8709&??.8709&??.8709\cr
??14000&??.9401&??.9401&??.9401&??.9401&??.9401&??.9401&??.9401&??.9401&??.9401\cr
??16000&?1.0193&?1.0193&?1.0193&?1.0193&?1.0193&?1.0193&?1.0193&?1.0193&?1.0193\cr
??18000&?1.1013&?1.1013&?1.1013&?1.1013&?1.1013&?1.1013&?1.1013&?1.1013&?1.1013\cr
??20000&?1.1812&?1.1812&?1.1812&?1.1812&?1.1812&?1.1812&?1.1812&?1.1812&?1.1812\cr
??23000&?1.2924&?1.2923&?1.2923&?1.2923&?1.2923&?1.2923&?1.2923&?1.2923&?1.2923\cr
??26000&?1.3924&?1.3912&?1.3910&?1.3909&?1.3909&?1.3909&?1.3909&?1.3909&?1.3909\cr
??30000&?1.5382&?1.5110&?1.5054&?1.5045&?1.5043&?1.5043&?1.5043&?1.5043&?1.5043\cr
??35000&?2.0191&?1.7321&?1.6428&?1.6268&?1.6236&?1.6230&?1.6229&?1.6228&?1.6228\cr
??40000&?2.8838&?2.2692&?1.8725&?1.7563&?1.7293&?1.7244&?1.7234&?1.7231&?1.7231\cr
??50000&?4.8825&?3.7789&?2.8867&?2.3089&?2.0008&?1.9172&?1.8978&?1.8937&?1.8928\cr
??55000&?6.0263&?4.6774&?3.5030&?2.7413&?2.2318&?2.0372&?1.9846&?1.9734&?1.9709\cr
??60000&?7.1605&?5.6211&?4.1928&?3.2122&?2.5295&?2.1919&?2.0796&?2.0541&?2.0485\cr
??65000&?8.3485&?6.5641&?4.9319&?3.7250&?2.8702&?2.3851&?2.1878&?2.1385&?2.1277\cr
??70000&?9.5500&?7.5398&?5.6795&?4.2795&?3.2435&?2.6112&?2.3122&?2.2285&?2.2100\cr
??75000&10.7275&?8.5356&?6.4405&?4.8545&?3.6478&?2.8634&?2.4534&?2.3253&?2.2963\cr
??80000&11.9105&?9.5197&?7.2226&?5.4379&?4.0762&?3.1376&?2.6104&?2.4292&?2.3872\cr
??85000&13.0986&10.4993&?8.0111&?6.0324&?4.5190&?3.4316&?2.7817&?2.5404&?2.4830\cr
??90000&14.2644&11.4852&?8.7950&?6.6392&?4.9713&?3.7422&?2.9663&?2.6588&?2.5838\cr
??95000&15.4087&12.4659&?9.5781&?7.2521&?5.4327&?4.0657&?3.1630&?2.7842&?2.6898\cr
?100000&16.5444&13.4307&10.3642&?7.8658&?5.9029&?4.3994&?3.3709&?2.9164&?2.8009\cr
?110000&18.7507&15.3297&11.9257&?9.0961&?6.8602&?5.0922&?3.8156&?3.2007&?3.0392\cr
?120000&20.8490&17.1802&13.4597&10.3301&?7.8285&?5.8128&?4.2928&?3.5104&?3.2987\cr
?130000&22.9556&18.9711&14.9712&11.5536&?8.8050&?6.5517&?4.7965&?3.8434&?3.5788\cr
?140000&25.0807&20.7632&16.4512&12.7658&?9.7829&?7.3023&?5.3206&?4.1971&?3.8777\cr
?150000&27.2496&22.5661&17.9197&13.9647&10.7566&?8.0593&?5.8588&?4.5677&?4.1925\cr
?160000&29.4830&24.3955&19.3919&15.1508&11.7248&?8.8172&?6.4055&?4.9509&?4.5196\cr
?170000&31.6978&26.2770&20.8733&16.3328&12.6870&?9.5716&?6.9562&?5.3423&?4.8550\cr
?180000&33.7774&28.1713&22.3837&17.5168&13.6449&10.3208&?7.5074&?5.7380&?5.1951\cr
?190000&35.6762&29.9913&23.9277&18.7097&14.6030&11.0649&?8.0562&?6.1350&?5.5370\cr
?200000&37.3981&31.6822&25.4660&19.9211&15.5653&11.8052&?8.6013&?6.5312&?5.8786\cr
?220000&40.3846&34.6508&28.3275&22.3775&17.5068&13.2836&?9.6784&?7.3161&?6.5547\cr
?230000&41.6858&35.9496&29.6102&23.5702&18.4789&14.0256&10.2115&?7.7035&?6.8878\cr
?240000&42.8797&37.1423&30.7949&24.7054&19.4410&14.7699&10.7423&?8.0874&?7.2170\cr
?250000&43.9790&38.2410&31.8893&25.7725&20.3818&15.5143&11.2714&?8.4679&?7.5423\cr
?265000&45.4738&39.7353&33.3807&27.2432&21.7311&16.6222&12.0631&?9.0329&?8.0230\cr
?280000&46.8098&41.0712&34.7152&28.5683&22.9870&17.7032&12.8514&?9.5917&?8.4957\cr
?300000&48.3852&42.6465&36.2898&30.1371&24.5059&19.0727&13.8900&10.3277&?9.1143\cr
?325000&50.0843&44.3456&37.9885&31.8330&26.1703&20.6348&15.1470&11.2310&?9.8692\cr
?350000&51.5429&45.8042&39.4471&33.2904&27.6108&22.0214&16.3308&12.1085&10.6025\cr
?375000&52.8089&47.0702&40.7131&34.5559&28.8658&23.2457&17.4230&12.9507&11.3103\cr
?400000&53.9183&48.1796&41.8225&35.6651&29.9677&24.3286&18.4181&13.7487&11.9882\cr
?425000&54.8985&49.1599&42.8028&36.6453&30.9424&25.2905&19.3199&14.4973&12.6323\cr
?450000&55.7710&50.0324&43.6754&37.5178&31.8106&26.1496&20.1359&15.1940&13.2401\cr
?475000&56.5528&50.8141&44.4572&38.2996&32.5888&26.9209&20.8752&15.8397&13.8104\cr
?500000&57.2572&51.5186&45.1617&39.0041&33.2903&27.6170&21.5466&16.4365&14.3438\cr
?525000&57.8954&52.1568&45.8000&39.6424&33.9259&28.2483&22.1583&16.9877&14.8414\cr
?550000&58.4763&52.7378&46.3809&40.2233&34.5045&28.8233&22.7174&17.4970&15.3051\cr
?600000&59.4947&53.7562&47.3994&41.2419&35.5191&29.8322&23.7017&18.4045&16.1400\cr
?650000&60.3585&54.6199&48.2633&42.1058&36.3797&30.6884&24.5398&19.1861&16.8668\cr
?700000&61.1004&55.3619&49.0053&42.8479&37.1191&31.4243&25.2616&19.8645&17.5026\cr
?750000&61.7448&56.0064&49.6498&43.4924&37.7613&32.0636&25.8896&20.4580&18.0620\cr
?800000&62.3098&56.5713&50.2148&44.0574&38.3243&32.6242&26.4409&20.9812&18.5573\cr
?850000&62.8092&57.0708&50.7143&44.5569&38.8221&33.1199&26.9287&21.4455&18.9984\cr
?900000&63.2539&57.5155&51.1590&45.0017&39.2653&33.5613&27.3633&21.8603&19.3935\cr
?950000&63.6524&57.9140&51.5576&45.4003&39.6625&33.9569&27.7531&22.2330&19.7493\cr
1000000&64.0117&58.2733&51.9168&45.7596&40.0206&34.3136&28.1046&22.5696&20.0712\cr
\noalign{\medskip \hrule}  }

\par\vfill\end